\documentclass[aps,prl,tighten,superscriptaddress,notitlepage]{revtex4-1}

\usepackage{graphicx}
\usepackage{hyperref}

\begin{document}

\title{Search for Dark Matter Subhalos in the High-Energy Gamma-ray Band with {\it Fermi} and the Cherenkov Telescope Array}

\author{D. Nieto}
\email{nieto@nevis.columbia.edu}
\affiliation{Physics Department, Columbia University, New York, NY 10027, USA}
\author{M. Errando}
\affiliation{Department of Physics and Astronomy, Barnard College, Columbia University, NY 10027, USA}
\author{L. Fortson}
\affiliation{School of Physics and Astronomy, University of Minnesota, Minneapolis, MN 55455, USA}
\author{B. Humensky}
\affiliation{Physics Department, Columbia University, New York, NY 10027, USA}
\author{R. Mukherjee}
\affiliation{Department of Physics and Astronomy, Barnard College, Columbia University, NY 10027, USA}
\author{M. A. S\'{a}nchez-Conde}
\affiliation{KIPAC/SLAC National Accelerator Laboratory, Menlo Park, CA 94025, USA}
\author{A. Smith}
\affiliation{Department of Physics and Astronomy, University of Utah, Salt Lake City, UT 84112, USA}
\author{A. Weinstein}
\affiliation{Department of Physics and Astronomy, Iowa State University, Ames, IA 50011, USA}
\author{M. Wood}
\affiliation{KIPAC/SLAC National Accelerator Laboratory, Menlo Park, CA 94025, USA}

\date{\today}

\begin{abstract}
We discuss the potential for the detection of dark matter and the
characterization of its particle nature via the observation of dark
matter subhalos. Specifically, we discuss the search for dark matter
Galactic subhalos in the gamma-ray band with the Large Area Telescope
on-board the {\it Fermi} gamma-ray Space Telescope, and the future
generation of imaging atmospheric Cherenkov telescopes, best
represented by the planned Cherenkov Telescope Array.
\end{abstract}

\maketitle

\section{I\lowercase{ntroduction}}

The concordance cosmological model, thoroughly validated by
observations, requires 83\% of the total mass density in the Universe
to be non-baryonic \cite{Komatsu:2011a}. The nature of the so-called
dark matter (DM) is one of the major open questions currently in
Physics, and on consideration that many plausible theories involve
new, exotic particles, the answer may
come from the confluence of Particle Physics and Astrophysics. Weakly
interacting massive particles (WIMPs) with masses in the GeV-TeV range
are well motivated DM particle candidates. WIMPs could self annihilate
or decay into Standard Model particles, and thus the nature of DM
could be revealed by the detection of these by-products, photons
amongst them~\cite{ParticleDarkMatter:2010a}.

A gamma-ray signal from WIMP annihilation would have a very
distinctive spectral shape: features such as annihilation lines
\cite{Bertone:2009a}, internal bremsstrahlung \cite{Bringmann:2008a},
as well as a characteristic cut-off at the DM particle mass, are
expected. The spectrum of WIMP annihilation or decay must be
universal, so even if one can detect all the previously mentioned
features in a single measured spectrum, an ultimate confirmation of
the DM origin of the signal would be the detection of the same
spectral shape in different gamma-ray
sources~\cite{Pieri:2009a,Lee:2009a,Ando:2008a,Baltz:2000a}.

Regions where high DM density is foreseen are the best candidates for
detection in the gamma-ray light of DM annihilation, since the
expected flux is proportional to the square of the DM density
integrated along the line of sight.  No clear DM signal has been
detected so far in any of the most promising targets, including dwarf
spheroidal galaxies
\cite{Aleksic:2011a,2012PhRvD..85f2001A,2011APh....34..608H,Aharonian:2008a},
the Galactic Center region
\cite{PhysRevLett.97.221102,2011PhRvL.106p1301A}, or galaxy clusters
\cite{Ackermann:2010ab,Aleksic:2010a,2012ApJ...750..123A}. Yet, other
regions of high DM density potentially exist in the Galaxy: N-body
cosmological simulations have uncovered how the cold DM distribution
evolves from almost homogeneous initial conditions into a hierarchical
and highly clustered state at
present~\cite{2012MNRAS.426.2046A,2012MNRAS.423.3018P}. High
resolution simulations of Milky Way-like DM halos indicate that the
halos should not be smooth but must exhibit a wealth of substructure
down to even the smallest scales resolved in the simulations
\cite{Diemand:2008a,Springel:2008a,Stadel:2008a}. These subhalos could
be too small to have accumulated enough baryonic matter to start star
formation and would therefore be essentially invisible to astronomical
observations~\cite{2005Natur.433..389D,2013MNRAS.431.1366S}. Some of
these subhalos could be massive enough and close enough that they
could be observed as bright gamma-ray emitters due to the annihilation
of DM particles \cite{Pieri:2008a}.  Since gamma-ray emission from DM
annihilation or decay is expected to be steady from any given subhalo,
such hypothetical sources would be found in deep sky
surveys~\cite{Kamionkowski:2010a}, and most likely would be among the
{\it Fermi}-Large Area Telescope (LAT) detected sources as
unassociated sources with no conventional counterpart at any other
wavelength. As already mentioned, the smoking gun for DM detection
could be a very distinct cut-off close to the DM particle mass. These
subhalos could be detected by {\it Fermi}-LAT, although the
distinctive cut-off would most likely be located at too high an energy
(see, e.g. the \emph{neutralino} mass lower limit of
$>46$~GeV~\cite{2012PhRvD..86a0001B}) to be detected within a
reasonable time (if at all). Therefore the complementarity between
{\it Fermi}-LAT and imaging atmospheric Cherenkov telescopes (IACTs)
emerges naturally. In the following, we assume the DM to be composed
of WIMPs with masses larger than the previously mentioned mass lower
limit, in such a way that the DM annihilation spectral cut-off would
lie in the very high energy (VHE, $>50$ GeV) gamma-ray band.

Searches for DM subhalos in the {\it Fermi}-LAT data have already been
conducted by the LAT team based on the first year of sky survey data,
looking for spatially extended, unassociated
sources~\cite{2012ApJ...747..121A}. Searches for DM subhalo candidates
in both {\it Fermi}-LAT First and Second Source Catalogs (1FGL and
2FGL catalog, respectively) have also been
presented~\cite{Nieto:2011c,Zechlin:2012b}, triggering IACT follow-up
observations~\cite{Nieto:2011e,2013arXiv1303.1406G}. Additionally, the
feasibility of DM subhalo searches with wide-field IACTs has been
studied~\cite{2011PhRvD..83a5003B}.

\section{U\lowercase{nassociated High-Energy Gamma-ray Sources as} DM \lowercase{subhalo candidates}}

The LAT, on board the {\it Fermi} Gamma-ray Space Telescope, has
detected a large number of Galactic and extragalactic point sources of
high-energy gamma rays above 100 MeV. The 2FGL catalog lists 1873 high
energy sources of gamma rays corresponding to 24 months of scientific
data taking~\cite{Nolan:2012a}. A large fraction of the high latitude
sources are associated with known sources, such as blazars. The
sources at low Galactic latitudes are largely pulsars, or other
supernova products. However, roughly $\sim400$ of the 1873 high
energy sources listed in the 2FGL catalog are
still unassociated~\cite{Ferrara:2012a}. The 2FGL catalog lists source
positions with good angular resolution, and flux and spectral
measurements for these sources, but none of them have clear
counterpart associations at other wavelengths. Thus, these sources
remain unassociated and are termed \emph{unassociated Fermi objects}
(UFOs).

It is quite likely that many of the UFOs will be identified with known
source types in the future with further observations. At the moment,
the error boxes of the sources are still large enough to possibly
harbor several likely candidates. While a systematic counterpart
search of the $\sim400$ UFOs individually is clearly unfeasible, it
may be possible to characterize the sources that are the best dark
matter (DM) subhalo candidates, and use the gamma-ray measurements and
observations at other wavebands to identify these targets as
astrophysical sources. Alternatively, ruling out a standard
astrophysical explanation for the gamma-ray emission would eventually
identify DM subhalos and lead to claims of indirect DM detection. In
order to do this in the most efficient way, one has to select the most
likely candidates for DM subhalos in the list of UFOs, and carry out
follow up observations at other frequencies, to look for an
astrophysical explanation.

DM subhalos are expected to be faint gamma-ray sources with apparent
sizes comparable to the point spread function of IACTs such as
VERITAS, H.E.S.S. or MAGIC. The best UFO candidates are those that
best resemble a gamma-ray signal originating from the annihilation of
DM particles in subhalos. Among the 2FGL
candidates, the best possible targets for DM subhalo searches are
likely to satisfy the following criteria: (a) location at high
Galactic latitudes, (b) steady gamma-ray flux, (c) hard spectrum and
(d) no obvious counterpart. 

The expected gamma-ray flux due to DM annihilation can be factored
into two terms: the so-called astrophysical and particle physics
factors. The latter factor is universal, and only depends on the DM
particle model. On the contrary, the astrophysical factor is
proportional to the DM density squared integrated along the line of
sight, and is thus source dependent. When trying to place upper limits
on the DM annihilation cross section, the uncertainty in the range of
astrophysical factor values one could expect from the population of DM
subhalos is a major drawback. The number of subhalos as a function of
their masses (known as subhalo mass function), as well as their
galactic radial distribution, and inner structures, vary from one
simulation to another. Consequently, any upper limit that could be
derived from the absence of DM subhalos in the gamma-ray sky would not
be model independent and would thus require certain assumptions on the
former parameters (see, e.g., \cite{2011PhRvD..83a5003B}). On the
other hand, this very same uncertainty is what makes these objects so
appealing, since there is the possibility of finding subhalos with
astrophysical factors large enough to produce detectable gamma-ray
fluxes within the sensitivity range of present and future
IACTs. Recent work~\cite{Klein:2012a} suggests that in Milky Way-like
galaxies a number of the subhalos should have astrophysical factors
comparable to, or larger than, those of the Milky Way dwarf galaxies
(often interpreted as the largest halo substructures). Consequently,
deep observations with IACTs on promising unassociated LAT sources
must be encouraged.

\section{P\lowercase{rospective Studies with} CTA}

\begin{figure}[!h]
   \centering
   \includegraphics[width=0.5\textwidth]{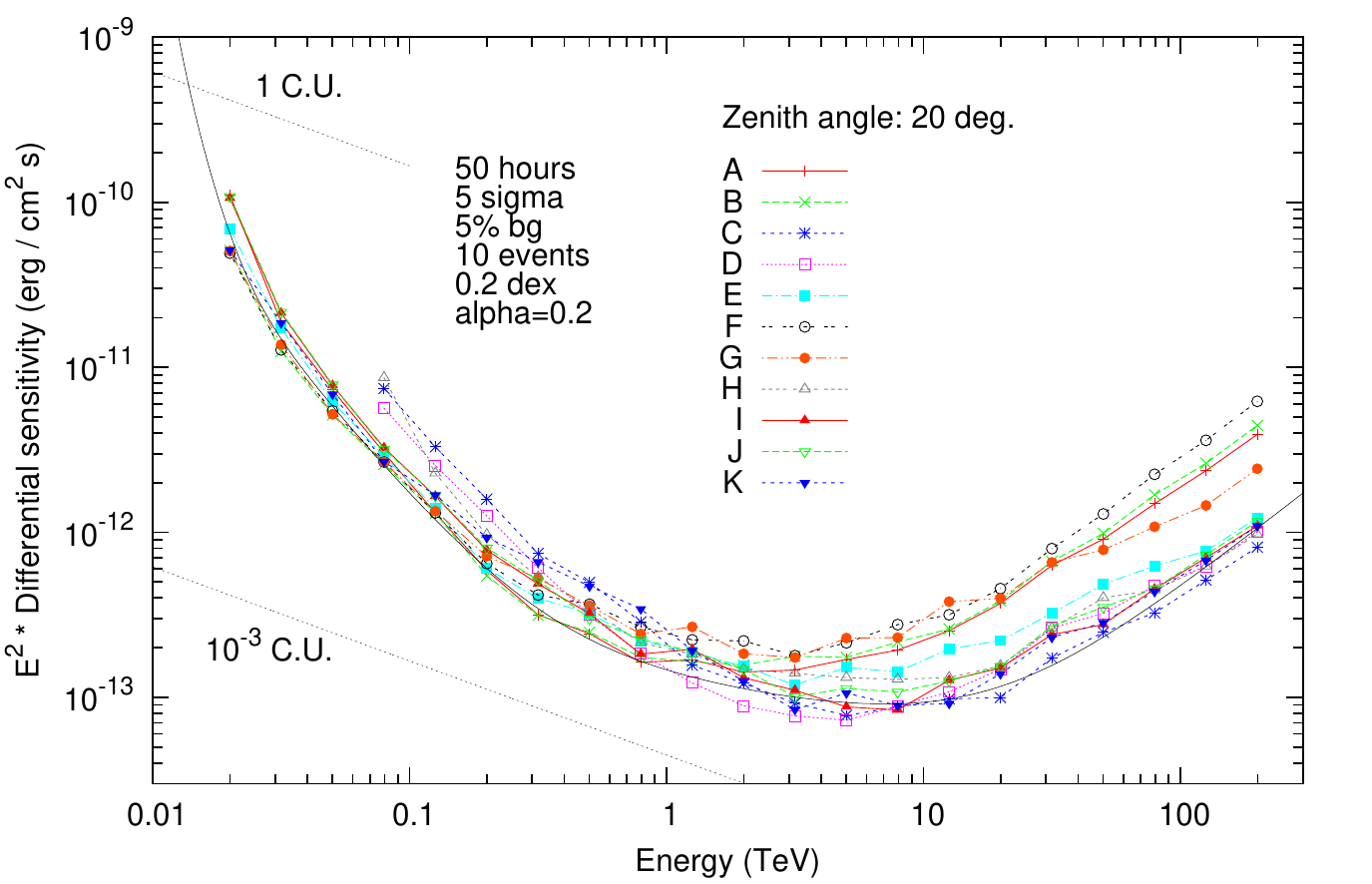}%
   \caption{\label{fig:sens} Expected point source differential
     sensitivity for 50 hours observation time at 20$^\circ$ zenith
     angle. Eleven different array layouts for CTA of similar costs,
     as described in~\cite{2013APh....43..171B}, are considered. The
     solid black line depicts an approximation of the best performance
     of any of these arrays at any energy. The sensitivity is provided
     in Crab Units (C.U.), where
     1~C.U.$=2.79\times10^{-7}\times($E$/$TeV$)^{-2.57}$~m$^{-2}$s$^{-1}$TeV$^{-1}$. Figure
     extracted from~\cite{2013APh....43..171B}.}
 \end{figure}

The next generation of IACTs is best represented by the planned
Cherenkov Telescope Array~\cite{CTAWeb:2011}. The expected performance
of the instrument has been extensively studied through detailed Monte
Carlo simulations~\cite{2013APh....43..171B} (the point source
sensitivity of the instrument is shown in Fig.~\ref{fig:sens} for
reference). The capabilities of CTA for DM detection have been
presented in~\cite{2013APh....43..189D}, where the detection prospects
for the most promising targets are studied.

\begin{figure}[!h]
   \centering
   \includegraphics[width=0.45\textwidth]{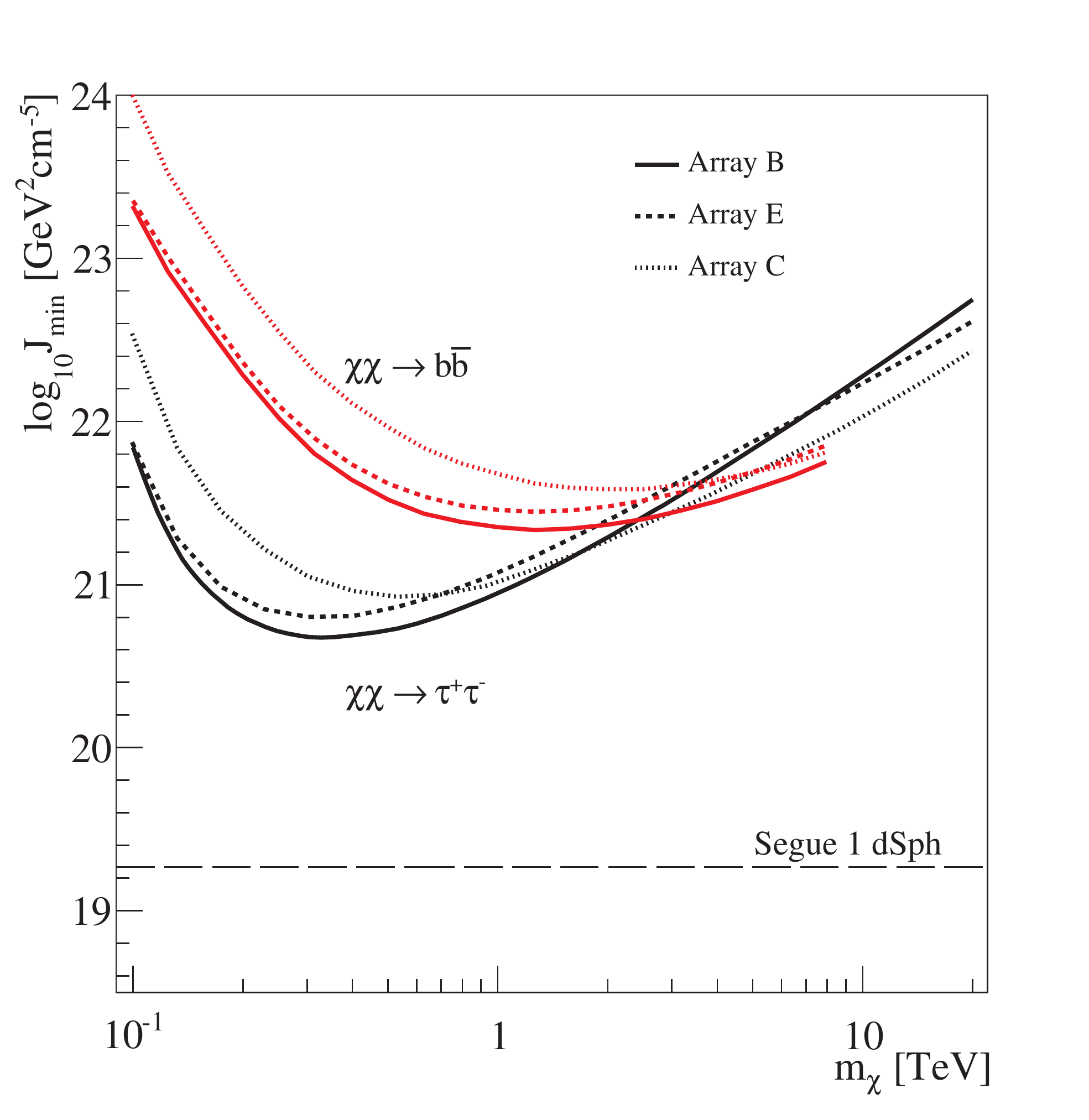}%
   \caption{\label{fig:jmins} Minimum value of the astrophysical
     factor required for a 5$\sigma$ detection with CTA as a function
     of the DM particle mass. An observation time T$_{obs}=100$ hours
     and an observation zenith angle of 20$^\circ$ are assumed. The
     results are computed for DM particles annihilating into $b
     \bar{b}$ (red lines) and $\tau^+ \tau^-$ (black lines) and
     assuming a canonical annihilation cross-section $<\sigma
     v>=3\times10^{-26}$ cm$^3$ s$^{-1}$. The sensitivities of three
     different array layouts shown in Fig.~\ref{fig:sens} have been
     considered (B, C and E). The astrophysical factor for Segue 1 is
     shown for reference. Figure extracted
     from~\cite{2013APh....43..189D}.}
 \end{figure}

Fig.~\ref{fig:jmins} illustrates the expected sensitivity of CTA to
generic, point-like sources showing DM annihilation-like spectral
shapes. Consequently, such prediction applies to any DM subhalo that
could be considered a point-like source for CTA. The sensitivity is
provided in terms of the minimum value of the astrophysical factor
required for a 5$\sigma$ detection after 100 hours of observations.
The astrophysical factor of the dwarf spheroidal galaxy Segue 1, the
highest among this type of objects, is shown for reference. Two
possible annihilation channels are shown: annihilation to $b\bar{b}$
quarks, as an example of a {\it soft} photon spectrum~\footnote{In the
  context of DM annihilation, the softest photon spectra could be
  found in TeV-mass WIMPs annihilating exclusively to {\it b} quarks,
  whose spectral shape could be grossly approximated by a power-law of
  spectral index ~3 (for reference purposes). For GeV WIMP masses,
  that index is closer to 2. Therefore, even if we use the adjective
  {\it soft} here, it does not necessarily have the same meaning as in
  conventional VHE sources, where a spectral index of 2 is considered
  hard.}, and annihilation to $\tau^+\tau^-$ leptons, as an example of
a {\it hard} photon spectrum. The projected sensitivity relies on the
performance inferred from the aforementioned Monte Carlo studies,
which do not take into consideration the US contribution to the array
elements. The inclusion of the US medium-size telescopes, most likely
using the novel Schwarzschild-Couder design, would provide a
significant sensitivity improvement of the instrument in the energy
range from 100 GeV to 10 TeV, a region most sensitive to DM searches
(as shown in Fig.~\ref{fig:jmins}). Such an effect can be quantified
as a factor of 2 to 3 improvement over the mentioned energy range
\cite{2012AIPC.1505..765J}, consequently boosting the capability of
the instrument for DM searches. A more detailed study of this
improvement can be found in an accompanying White
Paper~\cite{2013arXiv1305.0302W}.

\newpage
\begin{acknowledgments}
We acknowledge receiving helpful comments from Seth Digel and Emmanuel Moulin.
\end{acknowledgments}

\bibliography{references}

\begin{thebibliography}{42}%
\makeatletter
\providecommand \@ifxundefined [1]{%
 \@ifx{#1\undefined}
}%
\providecommand \@ifnum [1]{%
 \ifnum #1\expandafter \@firstoftwo
 \else \expandafter \@secondoftwo
 \fi
}%
\providecommand \@ifx [1]{%
 \ifx #1\expandafter \@firstoftwo
 \else \expandafter \@secondoftwo
 \fi
}%
\providecommand \natexlab [1]{#1}%
\providecommand \enquote  [1]{``#1''}%
\providecommand \bibnamefont  [1]{#1}%
\providecommand \bibfnamefont [1]{#1}%
\providecommand \citenamefont [1]{#1}%
\providecommand \href@noop [0]{\@secondoftwo}%
\providecommand \href [0]{\begingroup \@sanitize@url \@href}%
\providecommand \@href[1]{\@@startlink{#1}\@@href}%
\providecommand \@@href[1]{\endgroup#1\@@endlink}%
\providecommand \@sanitize@url [0]{\catcode `\\12\catcode `\$12\catcode
  `\&12\catcode `\#12\catcode `\^12\catcode `\_12\catcode `\%12\relax}%
\providecommand \@@startlink[1]{}%
\providecommand \@@endlink[0]{}%
\providecommand \url  [0]{\begingroup\@sanitize@url \@url }%
\providecommand \@url [1]{\endgroup\@href {#1}{\urlprefix }}%
\providecommand \urlprefix  [0]{URL }%
\providecommand \Eprint [0]{\href }%
\providecommand \doibase [0]{http://dx.doi.org/}%
\providecommand \selectlanguage [0]{\@gobble}%
\providecommand \bibinfo  [0]{\@secondoftwo}%
\providecommand \bibfield  [0]{\@secondoftwo}%
\providecommand \translation [1]{[#1]}%
\providecommand \BibitemOpen [0]{}%
\providecommand \bibitemStop [0]{}%
\providecommand \bibitemNoStop [0]{.\EOS\space}%
\providecommand \EOS [0]{\spacefactor3000\relax}%
\providecommand \BibitemShut  [1]{\csname bibitem#1\endcsname}%
\let\auto@bib@innerbib\@empty
\bibitem [{\citenamefont {{Komatsu}}\ \emph {et~al.}(2011)\citenamefont
  {{Komatsu}} \emph {et~al.}}]{Komatsu:2011a}%
  \BibitemOpen
  \bibfield  {author} {\bibinfo {author} {\bibfnamefont {E.}~\bibnamefont
  {{Komatsu}}} \emph {et~al.},\ }\href {\doibase 10.1088/0067-0049/192/2/18}
  {\bibfield  {journal} {\bibinfo  {journal} {ApJS}\ }\textbf {\bibinfo
  {volume} {192}},\ \bibinfo {pages} {18} (\bibinfo {year} {2011})},\ \Eprint
  {http://arxiv.org/abs/1001.4538} {arXiv:1001.4538 [astro-ph.CO]} \BibitemShut
  {NoStop}%
\bibitem [{\citenamefont {Gianfranco}(2010)}]{ParticleDarkMatter:2010a}%
  \BibitemOpen
  \bibinfo {editor} {\bibfnamefont {B.}~\bibnamefont {Gianfranco}},\ ed.,\
  \href@noop {} {\emph {\bibinfo {title} {{Particle Dark Matter}}}},\ \bibinfo
  {edition} {first published}\ ed.\ (\bibinfo  {publisher} {{Cambridge
  University Press}},\ \bibinfo {year} {2010})\BibitemShut {NoStop}%
\bibitem [{\citenamefont {Bertone}\ \emph {et~al.}(2009)\citenamefont
  {Bertone}, \citenamefont {Jackson}, \citenamefont {Shaughnessy},
  \citenamefont {Tait},\ and\ \citenamefont {Vallinotto}}]{Bertone:2009a}%
  \BibitemOpen
  \bibfield  {author} {\bibinfo {author} {\bibfnamefont {G.}~\bibnamefont
  {Bertone}}, \bibinfo {author} {\bibfnamefont {C.}~\bibnamefont {Jackson}},
  \bibinfo {author} {\bibfnamefont {G.}~\bibnamefont {Shaughnessy}}, \bibinfo
  {author} {\bibfnamefont {T.~M.}\ \bibnamefont {Tait}}, \ and\ \bibinfo
  {author} {\bibfnamefont {A.}~\bibnamefont {Vallinotto}},\ }\href {\doibase
  10.1103/PhysRevD.80.023512} {\bibfield  {journal} {\bibinfo  {journal}
  {Phys.Rev.D}\ }\textbf {\bibinfo {volume} {80}},\ \bibinfo {pages} {023512}
  (\bibinfo {year} {2009})},\ \Eprint
  {http://arxiv.org/abs/astro-ph.HE/0904.1442} {arXiv:astro-ph.HE/0904.1442
  [astro-ph.HE]} \BibitemShut {NoStop}%
\bibitem [{\citenamefont {Bringmann}\ \emph {et~al.}(2008)\citenamefont
  {Bringmann}, \citenamefont {Bergstrom},\ and\ \citenamefont
  {Edsjo}}]{Bringmann:2008a}%
  \BibitemOpen
  \bibfield  {author} {\bibinfo {author} {\bibfnamefont {T.}~\bibnamefont
  {Bringmann}}, \bibinfo {author} {\bibfnamefont {L.}~\bibnamefont
  {Bergstrom}}, \ and\ \bibinfo {author} {\bibfnamefont {J.}~\bibnamefont
  {Edsjo}},\ }\href {\doibase 10.1088/1126-6708/2008/01/049} {\bibfield
  {journal} {\bibinfo  {journal} {JHEP}\ }\textbf {\bibinfo {volume} {0801}},\
  \bibinfo {pages} {049} (\bibinfo {year} {2008})},\ \Eprint
  {http://arxiv.org/abs/0710.3169} {arXiv:0710.3169 [hep-ph]} \BibitemShut
  {NoStop}%
\bibitem [{\citenamefont {{Pieri}}\ \emph {et~al.}(2011)\citenamefont
  {{Pieri}}, \citenamefont {{Lavalle}}, \citenamefont {{Bertone}},\ and\
  \citenamefont {{Branchini}}}]{Pieri:2009a}%
  \BibitemOpen
  \bibfield  {author} {\bibinfo {author} {\bibfnamefont {L.}~\bibnamefont
  {{Pieri}}}, \bibinfo {author} {\bibfnamefont {J.}~\bibnamefont {{Lavalle}}},
  \bibinfo {author} {\bibfnamefont {G.}~\bibnamefont {{Bertone}}}, \ and\
  \bibinfo {author} {\bibfnamefont {E.}~\bibnamefont {{Branchini}}},\ }\href
  {\doibase 10.1103/PhysRevD.83.023518} {\bibfield  {journal} {\bibinfo
  {journal} {Phys.Rev.D}\ }\textbf {\bibinfo {volume} {83}},\ \bibinfo {pages}
  {023518} (\bibinfo {year} {2011})},\ \Eprint
  {http://arxiv.org/abs/astro-ph.HE/0908.0195} {arXiv:astro-ph.HE/0908.0195
  [astro-ph.HE]} \BibitemShut {NoStop}%
\bibitem [{\citenamefont {Lee}\ \emph {et~al.}(2009)\citenamefont {Lee},
  \citenamefont {Ando},\ and\ \citenamefont {Kamionkowski}}]{Lee:2009a}%
  \BibitemOpen
  \bibfield  {author} {\bibinfo {author} {\bibfnamefont {S.~K.}\ \bibnamefont
  {Lee}}, \bibinfo {author} {\bibfnamefont {S.}~\bibnamefont {Ando}}, \ and\
  \bibinfo {author} {\bibfnamefont {M.}~\bibnamefont {Kamionkowski}},\ }\href
  {\doibase 10.1088/1475-7516/2009/07/007} {\bibfield  {journal} {\bibinfo
  {journal} {JCAP}\ }\textbf {\bibinfo {volume} {0907}},\ \bibinfo {pages}
  {007} (\bibinfo {year} {2009})},\ \Eprint
  {http://arxiv.org/abs/astro-ph/0810.1284} {arXiv:astro-ph/0810.1284
  [astro-ph]} \BibitemShut {NoStop}%
\bibitem [{\citenamefont {Ando}\ \emph {et~al.}(2008)\citenamefont {Ando},
  \citenamefont {Kamionkowski}, \citenamefont {Lee},\ and\ \citenamefont
  {Koushiappas}}]{Ando:2008a}%
  \BibitemOpen
  \bibfield  {author} {\bibinfo {author} {\bibfnamefont {S.}~\bibnamefont
  {Ando}}, \bibinfo {author} {\bibfnamefont {M.}~\bibnamefont {Kamionkowski}},
  \bibinfo {author} {\bibfnamefont {S.~K.}\ \bibnamefont {Lee}}, \ and\
  \bibinfo {author} {\bibfnamefont {S.~M.}\ \bibnamefont {Koushiappas}},\
  }\href {\doibase 10.1103/PhysRevD.78.101301} {\bibfield  {journal} {\bibinfo
  {journal} {Phys.Rev.D}\ }\textbf {\bibinfo {volume} {78}},\ \bibinfo {pages}
  {101301} (\bibinfo {year} {2008})},\ \Eprint
  {http://arxiv.org/abs/astro-ph/0809.0886} {arXiv:astro-ph/0809.0886
  [astro-ph]} \BibitemShut {NoStop}%
\bibitem [{\citenamefont {Baltz}\ \emph {et~al.}(1999)\citenamefont {Baltz},
  \citenamefont {Briot}, \citenamefont {Salati}, \citenamefont {Taillet},\ and\
  \citenamefont {Silk}}]{Baltz:2000a}%
  \BibitemOpen
  \bibfield  {author} {\bibinfo {author} {\bibfnamefont {E.~A.}\ \bibnamefont
  {Baltz}}, \bibinfo {author} {\bibfnamefont {C.}~\bibnamefont {Briot}},
  \bibinfo {author} {\bibfnamefont {P.}~\bibnamefont {Salati}}, \bibinfo
  {author} {\bibfnamefont {R.}~\bibnamefont {Taillet}}, \ and\ \bibinfo
  {author} {\bibfnamefont {J.}~\bibnamefont {Silk}},\ }\href {\doibase
  10.1103/PhysRevD.61.023514} {\bibfield  {journal} {\bibinfo  {journal}
  {Phys.Rev.D}\ }\textbf {\bibinfo {volume} {61}},\ \bibinfo {pages} {023514}
  (\bibinfo {year} {1999})}\BibitemShut {NoStop}%
\bibitem [{\citenamefont {{Aleksi{\'c}}}\ \emph {et~al.}(2011)\citenamefont
  {{Aleksi{\'c}}} \emph {et~al.}}]{Aleksic:2011a}%
  \BibitemOpen
  \bibfield  {author} {\bibinfo {author} {\bibfnamefont {J.}~\bibnamefont
  {{Aleksi{\'c}}}} \emph {et~al.} (\bibinfo {collaboration} {MAGIC}),\ }\href
  {\doibase 10.1088/1475-7516/2011/06/035} {\bibfield  {journal} {\bibinfo
  {journal} {JCAP}\ }\textbf {\bibinfo {volume} {6}},\ \bibinfo {pages} {35}
  (\bibinfo {year} {2011})},\ \Eprint {http://arxiv.org/abs/1103.0477}
  {arXiv:1103.0477 [astro-ph.HE]} \BibitemShut {NoStop}%
\bibitem [{\citenamefont {{Aliu}}\ \emph {et~al.}(2012)\citenamefont {{Aliu}}
  \emph {et~al.}}]{2012PhRvD..85f2001A}%
  \BibitemOpen
  \bibfield  {author} {\bibinfo {author} {\bibfnamefont {E.}~\bibnamefont
  {{Aliu}}} \emph {et~al.} (\bibinfo {collaboration} {VERITAS}),\ }\href
  {\doibase 10.1103/PhysRevD.85.062001} {\bibfield  {journal} {\bibinfo
  {journal} {Phys.Rev.D}\ }\textbf {\bibinfo {volume} {85}},\ \bibinfo {eid}
  {062001} (\bibinfo {year} {2012})},\ \Eprint {http://arxiv.org/abs/1202.2144}
  {arXiv:1202.2144 [astro-ph.HE]} \BibitemShut {NoStop}%
\bibitem [{\citenamefont {{Abramowski}}\ \emph
  {et~al.}(2011{\natexlab{a}})\citenamefont {{Abramowski}} \emph
  {et~al.}}]{2011APh....34..608H}%
  \BibitemOpen
  \bibfield  {author} {\bibinfo {author} {\bibfnamefont {A.}~\bibnamefont
  {{Abramowski}}} \emph {et~al.} (\bibinfo {collaboration} {H.E.S.S}),\ }\href
  {\doibase 10.1016/j.astropartphys.2010.12.006} {\bibfield  {journal}
  {\bibinfo  {journal} {Astroparticle Physics}\ }\textbf {\bibinfo {volume}
  {34}},\ \bibinfo {pages} {608} (\bibinfo {year} {2011}{\natexlab{a}})},\
  \Eprint {http://arxiv.org/abs/1012.5602} {arXiv:1012.5602 [astro-ph.HE]}
  \BibitemShut {NoStop}%
\bibitem [{\citenamefont {Aharonian}\ \emph {et~al.}(2008)\citenamefont
  {Aharonian} \emph {et~al.}}]{Aharonian:2008a}%
  \BibitemOpen
  \bibfield  {author} {\bibinfo {author} {\bibfnamefont {F.}~\bibnamefont
  {Aharonian}} \emph {et~al.} (\bibinfo {collaboration} {H.E.S.S.}),\ }\href
  {\doibase 10.1016/j.astropartphys.2007.11.007} {\bibfield  {journal}
  {\bibinfo  {journal} {Astrop.Phys.}\ }\textbf {\bibinfo {volume} {29}},\
  \bibinfo {pages} {55} (\bibinfo {year} {2008})},\ \Eprint
  {http://arxiv.org/abs/astro-ph/0711.2369} {arXiv:astro-ph/0711.2369
  [astro-ph]} \BibitemShut {NoStop}%
\bibitem [{\citenamefont {Aharonian}\ \emph {et~al.}(2006)\citenamefont
  {Aharonian} \emph {et~al.}}]{PhysRevLett.97.221102}%
  \BibitemOpen
  \bibfield  {author} {\bibinfo {author} {\bibfnamefont {F.}~\bibnamefont
  {Aharonian}} \emph {et~al.} (\bibinfo {collaboration} {H.E.S.S.}),\ }\href
  {\doibase 10.1103/PhysRevLett.97.221102} {\bibfield  {journal} {\bibinfo
  {journal} {Phys. Rev. Lett.}\ }\textbf {\bibinfo {volume} {97}},\ \bibinfo
  {pages} {221102} (\bibinfo {year} {2006})}\BibitemShut {NoStop}%
\bibitem [{\citenamefont {{Abramowski}}\ \emph
  {et~al.}(2011{\natexlab{b}})\citenamefont {{Abramowski}} \emph
  {et~al.}}]{2011PhRvL.106p1301A}%
  \BibitemOpen
  \bibfield  {author} {\bibinfo {author} {\bibfnamefont {A.}~\bibnamefont
  {{Abramowski}}} \emph {et~al.} (\bibinfo {collaboration} {H.E.S.S.}),\ }\href
  {\doibase 10.1103/PhysRevLett.106.161301} {\bibfield  {journal} {\bibinfo
  {journal} {Phys.Rev.Lett.}\ }\textbf {\bibinfo {volume} {106}},\ \bibinfo
  {eid} {161301} (\bibinfo {year} {2011}{\natexlab{b}})},\ \Eprint
  {http://arxiv.org/abs/1103.3266} {arXiv:1103.3266 [astro-ph.HE]} \BibitemShut
  {NoStop}%
\bibitem [{\citenamefont {{Ackermann}}\ \emph {et~al.}(2010)\citenamefont
  {{Ackermann}} \emph {et~al.}}]{Ackermann:2010ab}%
  \BibitemOpen
  \bibfield  {author} {\bibinfo {author} {\bibfnamefont {M.}~\bibnamefont
  {{Ackermann}}} \emph {et~al.} (\bibinfo {collaboration} {Fermi-LAT}),\ }\href
  {\doibase 10.1088/1475-7516/2010/05/025} {\bibfield  {journal} {\bibinfo
  {journal} {JCAP}\ }\textbf {\bibinfo {volume} {5}},\ \bibinfo {pages} {25}
  (\bibinfo {year} {2010})},\ \Eprint {http://arxiv.org/abs/1002.2239}
  {arXiv:1002.2239 [astro-ph.CO]} \BibitemShut {NoStop}%
\bibitem [{\citenamefont {Aleksi\'c}\ \emph {et~al.}(2010)\citenamefont
  {Aleksi\'c} \emph {et~al.}}]{Aleksic:2010a}%
  \BibitemOpen
  \bibfield  {author} {\bibinfo {author} {\bibfnamefont {J.}~\bibnamefont
  {Aleksi\'c}} \emph {et~al.} (\bibinfo {collaboration} {MAGIC}),\ }\href
  {\doibase 10.1088/0004-637X/710/1/634} {\bibfield  {journal} {\bibinfo
  {journal} {ApJ}\ }\textbf {\bibinfo {volume} {710}},\ \bibinfo {pages} {634}
  (\bibinfo {year} {2010})},\ \Eprint {http://arxiv.org/abs/0909.3267}
  {arXiv:0909.3267 [astro-ph.HE]} \BibitemShut {NoStop}%
\bibitem [{\citenamefont {{Abramowski}}\ \emph {et~al.}(2012)\citenamefont
  {{Abramowski}} \emph {et~al.}}]{2012ApJ...750..123A}%
  \BibitemOpen
  \bibfield  {author} {\bibinfo {author} {\bibfnamefont {A.}~\bibnamefont
  {{Abramowski}}} \emph {et~al.} (\bibinfo {collaboration} {H.E.S.S.}),\ }\href
  {\doibase 10.1088/0004-637X/750/2/123} {\bibfield  {journal} {\bibinfo
  {journal} {ApJ}\ }\textbf {\bibinfo {volume} {750}},\ \bibinfo {eid} {123}
  (\bibinfo {year} {2012})},\ \Eprint {http://arxiv.org/abs/1202.5494}
  {arXiv:1202.5494 [astro-ph.HE]} \BibitemShut {NoStop}%
\bibitem [{\citenamefont {{Angulo}}\ \emph {et~al.}(2012)\citenamefont
  {{Angulo}}, \citenamefont {{Springel}}, \citenamefont {{White}},
  \citenamefont {{Jenkins}}, \citenamefont {{Baugh}},\ and\ \citenamefont
  {{Frenk}}}]{2012MNRAS.426.2046A}%
  \BibitemOpen
  \bibfield  {author} {\bibinfo {author} {\bibfnamefont {R.~E.}\ \bibnamefont
  {{Angulo}}}, \bibinfo {author} {\bibfnamefont {V.}~\bibnamefont
  {{Springel}}}, \bibinfo {author} {\bibfnamefont {S.~D.~M.}\ \bibnamefont
  {{White}}}, \bibinfo {author} {\bibfnamefont {A.}~\bibnamefont {{Jenkins}}},
  \bibinfo {author} {\bibfnamefont {C.~M.}\ \bibnamefont {{Baugh}}}, \ and\
  \bibinfo {author} {\bibfnamefont {C.~S.}\ \bibnamefont {{Frenk}}},\ }\href
  {\doibase 10.1111/j.1365-2966.2012.21830.x} {\bibfield  {journal} {\bibinfo
  {journal} {MNRAS}\ }\textbf {\bibinfo {volume} {426}},\ \bibinfo {pages}
  {2046} (\bibinfo {year} {2012})},\ \Eprint {http://arxiv.org/abs/1203.3216}
  {arXiv:1203.3216 [astro-ph.CO]} \BibitemShut {NoStop}%
\bibitem [{\citenamefont {{Prada}}\ \emph {et~al.}(2012)\citenamefont
  {{Prada}}, \citenamefont {{Klypin}}, \citenamefont {{Cuesta}}, \citenamefont
  {{Betancort-Rijo}},\ and\ \citenamefont {{Primack}}}]{2012MNRAS.423.3018P}%
  \BibitemOpen
  \bibfield  {author} {\bibinfo {author} {\bibfnamefont {F.}~\bibnamefont
  {{Prada}}}, \bibinfo {author} {\bibfnamefont {A.~A.}\ \bibnamefont
  {{Klypin}}}, \bibinfo {author} {\bibfnamefont {A.~J.}\ \bibnamefont
  {{Cuesta}}}, \bibinfo {author} {\bibfnamefont {J.~E.}\ \bibnamefont
  {{Betancort-Rijo}}}, \ and\ \bibinfo {author} {\bibfnamefont
  {J.}~\bibnamefont {{Primack}}},\ }\href {\doibase
  10.1111/j.1365-2966.2012.21007.x} {\bibfield  {journal} {\bibinfo  {journal}
  {MNRAS}\ }\textbf {\bibinfo {volume} {423}},\ \bibinfo {pages} {3018}
  (\bibinfo {year} {2012})},\ \Eprint {http://arxiv.org/abs/1104.5130}
  {arXiv:1104.5130 [astro-ph.CO]} \BibitemShut {NoStop}%
\bibitem [{\citenamefont {Diemand}\ \emph {et~al.}(2008)\citenamefont
  {Diemand}, \citenamefont {Kuhlen}, \citenamefont {Madau}, \citenamefont
  {Zemp}, \citenamefont {Moore} \emph {et~al.}}]{Diemand:2008a}%
  \BibitemOpen
  \bibfield  {author} {\bibinfo {author} {\bibfnamefont {J.}~\bibnamefont
  {Diemand}}, \bibinfo {author} {\bibfnamefont {M.}~\bibnamefont {Kuhlen}},
  \bibinfo {author} {\bibfnamefont {P.}~\bibnamefont {Madau}}, \bibinfo
  {author} {\bibfnamefont {M.}~\bibnamefont {Zemp}}, \bibinfo {author}
  {\bibfnamefont {B.}~\bibnamefont {Moore}},  \emph {et~al.},\ }\href {\doibase
  10.1038/nature07153} {\bibfield  {journal} {\bibinfo  {journal} {Nature}\
  }\textbf {\bibinfo {volume} {454}},\ \bibinfo {pages} {735} (\bibinfo {year}
  {2008})},\ \Eprint {http://arxiv.org/abs/0805.1244} {arXiv:0805.1244
  [astro-ph]} \BibitemShut {NoStop}%
\bibitem [{\citenamefont {Springel}\ \emph {et~al.}(2008)\citenamefont
  {Springel}, \citenamefont {White}, \citenamefont {Frenk}, \citenamefont
  {Navarro}, \citenamefont {Jenkins} \emph {et~al.}}]{Springel:2008a}%
  \BibitemOpen
  \bibfield  {author} {\bibinfo {author} {\bibfnamefont {V.}~\bibnamefont
  {Springel}}, \bibinfo {author} {\bibfnamefont {S.}~\bibnamefont {White}},
  \bibinfo {author} {\bibfnamefont {C.}~\bibnamefont {Frenk}}, \bibinfo
  {author} {\bibfnamefont {J.}~\bibnamefont {Navarro}}, \bibinfo {author}
  {\bibfnamefont {A.}~\bibnamefont {Jenkins}},  \emph {et~al.},\ }\href
  {\doibase 10.1038/nature07411} {\bibfield  {journal} {\bibinfo  {journal}
  {Nature}\ }\textbf {\bibinfo {volume} {456}},\ \bibinfo {pages} {73}
  (\bibinfo {year} {2008})}\BibitemShut {NoStop}%
\bibitem [{\citenamefont {Stadel}\ \emph {et~al.}(2009)\citenamefont {Stadel},
  \citenamefont {Potter}, \citenamefont {Moore}, \citenamefont {Diemand},
  \citenamefont {Madau} \emph {et~al.}}]{Stadel:2008a}%
  \BibitemOpen
  \bibfield  {author} {\bibinfo {author} {\bibfnamefont {J.}~\bibnamefont
  {Stadel}}, \bibinfo {author} {\bibfnamefont {D.}~\bibnamefont {Potter}},
  \bibinfo {author} {\bibfnamefont {B.}~\bibnamefont {Moore}}, \bibinfo
  {author} {\bibfnamefont {J.}~\bibnamefont {Diemand}}, \bibinfo {author}
  {\bibfnamefont {P.}~\bibnamefont {Madau}},  \emph {et~al.},\ }\href {\doibase
  10.1111/j.1745-3933.2009.00699.x} {\bibfield  {journal} {\bibinfo  {journal}
  {MNRAS}\ }\textbf {\bibinfo {volume} {398}},\ \bibinfo {pages} {L21}
  (\bibinfo {year} {2009})},\ \Eprint {http://arxiv.org/abs/0808.2981}
  {arXiv:0808.2981} \BibitemShut {NoStop}%
\bibitem [{\citenamefont {{Diemand}}\ \emph {et~al.}(2005)\citenamefont
  {{Diemand}}, \citenamefont {{Moore}},\ and\ \citenamefont
  {{Stadel}}}]{2005Natur.433..389D}%
  \BibitemOpen
  \bibfield  {author} {\bibinfo {author} {\bibfnamefont {J.}~\bibnamefont
  {{Diemand}}}, \bibinfo {author} {\bibfnamefont {B.}~\bibnamefont {{Moore}}},
  \ and\ \bibinfo {author} {\bibfnamefont {J.}~\bibnamefont {{Stadel}}},\
  }\href {\doibase 10.1038/nature03270} {\bibfield  {journal} {\bibinfo
  {journal} {Nature}\ }\textbf {\bibinfo {volume} {433}},\ \bibinfo {pages}
  {389} (\bibinfo {year} {2005})},\ \Eprint
  {http://arxiv.org/abs/astro-ph/0501589} {astro-ph/0501589} \BibitemShut
  {NoStop}%
\bibitem [{\citenamefont {{Sawala}}\ \emph {et~al.}(2013)\citenamefont
  {{Sawala}}, \citenamefont {{Frenk}}, \citenamefont {{Crain}}, \citenamefont
  {{Jenkins}}, \citenamefont {{Schaye}}, \citenamefont {{Theuns}},\ and\
  \citenamefont {{Zavala}}}]{2013MNRAS.431.1366S}%
  \BibitemOpen
  \bibfield  {author} {\bibinfo {author} {\bibfnamefont {T.}~\bibnamefont
  {{Sawala}}}, \bibinfo {author} {\bibfnamefont {C.~S.}\ \bibnamefont
  {{Frenk}}}, \bibinfo {author} {\bibfnamefont {R.~A.}\ \bibnamefont
  {{Crain}}}, \bibinfo {author} {\bibfnamefont {A.}~\bibnamefont {{Jenkins}}},
  \bibinfo {author} {\bibfnamefont {J.}~\bibnamefont {{Schaye}}}, \bibinfo
  {author} {\bibfnamefont {T.}~\bibnamefont {{Theuns}}}, \ and\ \bibinfo
  {author} {\bibfnamefont {J.}~\bibnamefont {{Zavala}}},\ }\href {\doibase
  10.1093/mnras/stt259} {\bibfield  {journal} {\bibinfo  {journal} {MNRAS}\
  }\textbf {\bibinfo {volume} {431}},\ \bibinfo {pages} {1366} (\bibinfo {year}
  {2013})},\ \Eprint {http://arxiv.org/abs/1206.6495} {arXiv:1206.6495
  [astro-ph.CO]} \BibitemShut {NoStop}%
\bibitem [{\citenamefont {Pieri}\ \emph {et~al.}(2008)\citenamefont {Pieri},
  \citenamefont {Bertone},\ and\ \citenamefont {Branchini}}]{Pieri:2008a}%
  \BibitemOpen
  \bibfield  {author} {\bibinfo {author} {\bibfnamefont {L.}~\bibnamefont
  {Pieri}}, \bibinfo {author} {\bibfnamefont {G.}~\bibnamefont {Bertone}}, \
  and\ \bibinfo {author} {\bibfnamefont {E.}~\bibnamefont {Branchini}},\ }\href
  {\doibase 10.1111/j.1365-2966.2007.12828.x} {\bibfield  {journal} {\bibinfo
  {journal} {MNRAS}\ }\textbf {\bibinfo {volume} {384}},\ \bibinfo {pages}
  {1627} (\bibinfo {year} {2008})},\ \Eprint {http://arxiv.org/abs/0706.2101}
  {arXiv:0706.2101 [astro-ph]} \BibitemShut {NoStop}%
\bibitem [{\citenamefont {Kamionkowski}\ \emph {et~al.}(2010)\citenamefont
  {Kamionkowski}, \citenamefont {Koushiappas},\ and\ \citenamefont
  {Kuhlen}}]{Kamionkowski:2010a}%
  \BibitemOpen
  \bibfield  {author} {\bibinfo {author} {\bibfnamefont {M.}~\bibnamefont
  {Kamionkowski}}, \bibinfo {author} {\bibfnamefont {S.~M.}\ \bibnamefont
  {Koushiappas}}, \ and\ \bibinfo {author} {\bibfnamefont {M.}~\bibnamefont
  {Kuhlen}},\ }\href {\doibase 10.1103/PhysRevD.81.043532} {\bibfield
  {journal} {\bibinfo  {journal} {Phys.Rev.D}\ }\textbf {\bibinfo {volume}
  {81}},\ \bibinfo {pages} {043532} (\bibinfo {year} {2010})},\ \Eprint
  {http://arxiv.org/abs/1001.3144} {arXiv:1001.3144 [astro-ph]} \BibitemShut
  {NoStop}%
\bibitem [{\citenamefont {{Beringer}}\ \emph {et~al.}(2012)\citenamefont
  {{Beringer}} \emph {et~al.}}]{2012PhRvD..86a0001B}%
  \BibitemOpen
  \bibfield  {author} {\bibinfo {author} {\bibfnamefont {J.}~\bibnamefont
  {{Beringer}}} \emph {et~al.},\ }\href {\doibase 10.1103/PhysRevD.86.010001}
  {\bibfield  {journal} {\bibinfo  {journal} {Phys.Rev.D}\ }\textbf {\bibinfo
  {volume} {86}},\ \bibinfo {eid} {010001} (\bibinfo {year}
  {2012})}\BibitemShut {NoStop}%
\bibitem [{\citenamefont {{Ackermann}}\ \emph {et~al.}(2012)\citenamefont
  {{Ackermann}} \emph {et~al.}}]{2012ApJ...747..121A}%
  \BibitemOpen
  \bibfield  {author} {\bibinfo {author} {\bibfnamefont {M.}~\bibnamefont
  {{Ackermann}}} \emph {et~al.} (\bibinfo {collaboration} {{Fermi-LAT}}),\
  }\href {\doibase 10.1088/0004-637X/747/2/121} {\bibfield  {journal} {\bibinfo
   {journal} {ApJ}\ }\textbf {\bibinfo {volume} {747}},\ \bibinfo {eid} {121}
  (\bibinfo {year} {2012})},\ \Eprint {http://arxiv.org/abs/1201.2691}
  {arXiv:1201.2691 [astro-ph.HE]} \BibitemShut {NoStop}%
\bibitem [{\citenamefont {{Nieto}}\ \emph
  {et~al.}(2011{\natexlab{a}})\citenamefont {{Nieto}}, \citenamefont
  {{Mart{\'{\i}}nez}}, \citenamefont {{Mirabal}}, \citenamefont {{Barrio}},
  \citenamefont {{Satalecka}}, \citenamefont {{Pardo}},\ and\ \citenamefont
  {{Lozano}}}]{Nieto:2011c}%
  \BibitemOpen
  \bibfield  {author} {\bibinfo {author} {\bibfnamefont {D.}~\bibnamefont
  {{Nieto}}}, \bibinfo {author} {\bibfnamefont {V.}~\bibnamefont
  {{Mart{\'{\i}}nez}}}, \bibinfo {author} {\bibfnamefont {N.}~\bibnamefont
  {{Mirabal}}}, \bibinfo {author} {\bibfnamefont {J.~A.}\ \bibnamefont
  {{Barrio}}}, \bibinfo {author} {\bibfnamefont {K.}~\bibnamefont
  {{Satalecka}}}, \bibinfo {author} {\bibfnamefont {S.}~\bibnamefont
  {{Pardo}}}, \ and\ \bibinfo {author} {\bibfnamefont {I.}~\bibnamefont
  {{Lozano}}},\ }\href@noop {} {\bibfield  {journal} {\bibinfo  {journal}
  {$3^{rd}$ Fermi Symposium, Rome}\ } (\bibinfo {year} {2011}{\natexlab{a}})},\
  \Eprint {http://arxiv.org/abs/1110.4744} {arXiv:1110.4744 [astro-ph.HE]}
  \BibitemShut {NoStop}%
\bibitem [{\citenamefont {{Zechlin}}\ and\ \citenamefont
  {{Horns}}(2012)}]{Zechlin:2012b}%
  \BibitemOpen
  \bibfield  {author} {\bibinfo {author} {\bibfnamefont {H.-S.}\ \bibnamefont
  {{Zechlin}}}\ and\ \bibinfo {author} {\bibfnamefont {D.}~\bibnamefont
  {{Horns}}},\ }\href {\doibase 10.1088/1475-7516/2012/11/050} {\bibfield
  {journal} {\bibinfo  {journal} {JCAP}\ }\textbf {\bibinfo {volume} {11}},\
  \bibinfo {eid} {050} (\bibinfo {year} {2012})},\ \Eprint
  {http://arxiv.org/abs/1210.3852} {arXiv:1210.3852 [astro-ph.HE]} \BibitemShut
  {NoStop}%
\bibitem [{\citenamefont {{Nieto}}\ \emph
  {et~al.}(2011{\natexlab{b}})\citenamefont {{Nieto}} \emph
  {et~al.}}]{Nieto:2011e}%
  \BibitemOpen
  \bibfield  {author} {\bibinfo {author} {\bibfnamefont {D.}~\bibnamefont
  {{Nieto}}} \emph {et~al.},\ }\href@noop {} {\bibfield  {journal} {\bibinfo
  {journal} {{32$^{nd}$ International Cosmic Ray Conference, Beijing}}\ }
  (\bibinfo {year} {2011}{\natexlab{b}})},\ \Eprint
  {http://arxiv.org/abs/1109.5935} {arXiv:1109.5935 [astro-ph.HE]} \BibitemShut
  {NoStop}%
\bibitem [{\citenamefont {{Geringer-Sameth}}\ and\ \citenamefont {{for the
  VERITAS Collaboration}}(2012)}]{2013arXiv1303.1406G}%
  \BibitemOpen
  \bibfield  {author} {\bibinfo {author} {\bibfnamefont {A.}~\bibnamefont
  {{Geringer-Sameth}}}\ and\ \bibinfo {author} {\bibnamefont {{for the VERITAS
  Collaboration}}},\ }\href@noop {} {\bibfield  {journal} {\bibinfo  {journal}
  {4$^{th}$ Fermi Symposium, Monterrey, CA}\ } (\bibinfo {year} {2012})},\
  \Eprint {http://arxiv.org/abs/1303.1406} {arXiv:1303.1406 [astro-ph.HE]}
  \BibitemShut {NoStop}%
\bibitem [{\citenamefont {{Brun}}\ \emph {et~al.}(2011)\citenamefont {{Brun}},
  \citenamefont {{Moulin}}, \citenamefont {{Diemand}},\ and\ \citenamefont
  {{Glicenstein}}}]{2011PhRvD..83a5003B}%
  \BibitemOpen
  \bibfield  {author} {\bibinfo {author} {\bibfnamefont {P.}~\bibnamefont
  {{Brun}}}, \bibinfo {author} {\bibfnamefont {E.}~\bibnamefont {{Moulin}}},
  \bibinfo {author} {\bibfnamefont {J.}~\bibnamefont {{Diemand}}}, \ and\
  \bibinfo {author} {\bibfnamefont {J.-F.}\ \bibnamefont {{Glicenstein}}},\
  }\href {\doibase 10.1103/PhysRevD.83.015003} {\bibfield  {journal} {\bibinfo
  {journal} {Phys.Rev.D}\ }\textbf {\bibinfo {volume} {83}},\ \bibinfo {eid}
  {015003} (\bibinfo {year} {2011})},\ \Eprint {http://arxiv.org/abs/1012.4766}
  {arXiv:1012.4766 [astro-ph.HE]} \BibitemShut {NoStop}%
\bibitem [{\citenamefont {{Nolan}}\ \emph {et~al.}(2012)\citenamefont
  {{Nolan}}, \citenamefont {{Abdo}} \emph {et~al.}}]{Nolan:2012a}%
  \BibitemOpen
  \bibfield  {author} {\bibinfo {author} {\bibfnamefont {P.~L.}\ \bibnamefont
  {{Nolan}}}, \bibinfo {author} {\bibfnamefont {A.~A.}\ \bibnamefont {{Abdo}}},
   \emph {et~al.} (\bibinfo {collaboration} {{Fermi-LAT}}),\ }\href {\doibase
  10.1088/0067-0049/199/2/31} {\bibfield  {journal} {\bibinfo  {journal}
  {ApJS}\ }\textbf {\bibinfo {volume} {199}},\ \bibinfo {eid} {31} (\bibinfo
  {year} {2012})},\ \Eprint {http://arxiv.org/abs/1108.1435} {arXiv:1108.1435
  [astro-ph.HE]} \BibitemShut {NoStop}%
\bibitem [{\citenamefont {Ferrara}\ \emph {et~al.}(2012)\citenamefont {Ferrara}
  \emph {et~al.}}]{Ferrara:2012a}%
  \BibitemOpen
  \bibfield  {author} {\bibinfo {author} {\bibfnamefont {E.}~\bibnamefont
  {Ferrara}} \emph {et~al.} (\bibinfo {collaboration} {Fermi-LAT}),\
  }\href@noop {} {\bibfield  {journal} {\bibinfo  {journal} {4$^{th}$~Fermi
  Symposium, Monterrey, CA}\ } (\bibinfo {year} {2012})}\BibitemShut {NoStop}%
\bibitem [{\citenamefont {Klein}\ \emph {et~al.}(2012)\citenamefont {Klein},
  \citenamefont {S{\'a}nchez-Conde}, \citenamefont {Drlica-Wagner},\ and\
  \citenamefont {Bloom}}]{Klein:2012a}%
  \BibitemOpen
  \bibfield  {author} {\bibinfo {author} {\bibfnamefont {A.}~\bibnamefont
  {Klein}}, \bibinfo {author} {\bibfnamefont {M.~A.}\ \bibnamefont
  {S{\'a}nchez-Conde}}, \bibinfo {author} {\bibfnamefont {A.}~\bibnamefont
  {Drlica-Wagner}}, \ and\ \bibinfo {author} {\bibfnamefont {E.}~\bibnamefont
  {Bloom}},\ }\href@noop {} {\bibfield  {journal} {\bibinfo  {journal}
  {4$^{th}$~Fermi Symposium, Monterrey, CA}\ } (\bibinfo {year}
  {2012})}\BibitemShut {NoStop}%
\bibitem [{\citenamefont {{CTA Consortium}}(2013)}]{CTAWeb:2011}%
  \BibitemOpen
  \bibfield  {author} {\bibinfo {author} {\bibnamefont {{CTA Consortium}}},\
  }\href@noop {} {\enquote {\bibinfo {title} {{Cherenkov Telescope Array
  Homepage}},}\ }\bibinfo {howpublished} {\url{www.cta-observatory.org}}
  (\bibinfo {year} {2013})\BibitemShut {NoStop}%
\bibitem [{\citenamefont {{Bernl{\"o}hr}}\ \emph {et~al.}(2013)\citenamefont
  {{Bernl{\"o}hr}} \emph {et~al.}}]{2013APh....43..171B}%
  \BibitemOpen
  \bibfield  {author} {\bibinfo {author} {\bibfnamefont {K.}~\bibnamefont
  {{Bernl{\"o}hr}}} \emph {et~al.},\ }\href {\doibase
  10.1016/j.astropartphys.2012.10.002} {\bibfield  {journal} {\bibinfo
  {journal} {Astroparticle Physics}\ }\textbf {\bibinfo {volume} {43}},\
  \bibinfo {pages} {171} (\bibinfo {year} {2013})},\ \Eprint
  {http://arxiv.org/abs/1210.3503} {arXiv:1210.3503 [astro-ph.IM]} \BibitemShut
  {NoStop}%
\bibitem [{\citenamefont {{Doro}}\ \emph {et~al.}(2013)\citenamefont {{Doro}}
  \emph {et~al.}}]{2013APh....43..189D}%
  \BibitemOpen
  \bibfield  {author} {\bibinfo {author} {\bibfnamefont {M.}~\bibnamefont
  {{Doro}}} \emph {et~al.},\ }\href {\doibase
  10.1016/j.astropartphys.2012.08.002} {\bibfield  {journal} {\bibinfo
  {journal} {Astroparticle Physics}\ }\textbf {\bibinfo {volume} {43}},\
  \bibinfo {pages} {189} (\bibinfo {year} {2013})},\ \Eprint
  {http://arxiv.org/abs/1208.5356} {arXiv:1208.5356 [astro-ph.IM]} \BibitemShut
  {NoStop}%
\bibitem [{Note1()}]{Note1}%
  \BibitemOpen
  \bibinfo {note} {In the context of DM annihilation, the softest photon
  spectra could be found in TeV-mass WIMPs annihilating exclusively to
  {\protect \it b} quarks, whose spectral shape could be grossly approximated
  by a power-law of spectral index ~3 (for reference purposes). For GeV WIMP
  masses, that index is closer to 2. Therefore, even if we use the adjective
  {\protect \it soft} here, it does not necessarily have the same meaning as in
  conventional VHE sources, where a spectral index of 2 is considered
  hard.}\BibitemShut {Stop}%
\bibitem [{\citenamefont {{Jogler}}\ \emph {et~al.}(2012)\citenamefont
  {{Jogler}}, \citenamefont {{Wood}},\ and\ \citenamefont {{Dumm for the CTA
  Consortium}}}]{2012AIPC.1505..765J}%
  \BibitemOpen
  \bibfield  {author} {\bibinfo {author} {\bibfnamefont {T.}~\bibnamefont
  {{Jogler}}}, \bibinfo {author} {\bibfnamefont {M.~D.}\ \bibnamefont
  {{Wood}}}, \ and\ \bibinfo {author} {\bibfnamefont {J.}~\bibnamefont {{Dumm
  for the CTA Consortium}}},\ }in\ \href {\doibase 10.1063/1.4772372} {\emph
  {\bibinfo {booktitle} {American Institute of Physics Conference Series}}},\
  \bibinfo {series} {American Institute of Physics Conference Series}, Vol.\
  \bibinfo {volume} {1505},\ \bibinfo {editor} {edited by\ \bibinfo {editor}
  {\bibfnamefont {F.~A.}\ \bibnamefont {{Aharonian}}}, \bibinfo {editor}
  {\bibfnamefont {W.}~\bibnamefont {{Hofmann}}}, \ and\ \bibinfo {editor}
  {\bibfnamefont {F.~M.}\ \bibnamefont {{Rieger}}}}\ (\bibinfo {year} {2012})\
  pp.\ \bibinfo {pages} {765--768},\ \Eprint {http://arxiv.org/abs/1211.3181}
  {arXiv:1211.3181 [astro-ph.IM]} \BibitemShut {NoStop}%
\bibitem [{\citenamefont {{Wood}}\ \emph {et~al.}(2013)\citenamefont {{Wood}},
  \citenamefont {{Buckley}}, \citenamefont {{Digel}}, \citenamefont {{Funk}},
  \citenamefont {{Nieto}},\ and\ \citenamefont
  {{Sanchez-Conde}}}]{2013arXiv1305.0302W}%
  \BibitemOpen
  \bibfield  {author} {\bibinfo {author} {\bibfnamefont {M.}~\bibnamefont
  {{Wood}}}, \bibinfo {author} {\bibfnamefont {J.}~\bibnamefont {{Buckley}}},
  \bibinfo {author} {\bibfnamefont {S.}~\bibnamefont {{Digel}}}, \bibinfo
  {author} {\bibfnamefont {S.}~\bibnamefont {{Funk}}}, \bibinfo {author}
  {\bibfnamefont {D.}~\bibnamefont {{Nieto}}}, \ and\ \bibinfo {author}
  {\bibfnamefont {M.~A.}\ \bibnamefont {{Sanchez-Conde}}},\ }\href@noop {}
  {\bibfield  {journal} {\bibinfo  {journal} {Snowmass 2013 Proceedings,
  SNOW13-00016}\ } (\bibinfo {year} {2013})},\ \Eprint
  {http://arxiv.org/abs/1305.0302} {arXiv:1305.0302 [astro-ph.HE]} \BibitemShut
  {NoStop}%
\end{thebibliography}%

\end{document}